# Magnetic order and phase transition in the iron oxysulfide $La_2O_2Fe_2OS_2$

Reeya K. Oogarah[1], Emmanuelle Suard[2] and Emma E. McCabe[1*]

[1] School of Physical Sciences, Ingram Building, University of Kent, Canterbury, Kent, CT2 7NH, U.K.
[2] Intitut Laue Langevin, 71 avenue des Martyrs - CS 20156 - 38042 GRENOBLE CEDEX 9, France
* corresponding author: e.e.mccabe@kent.ac.uk

**Abstract**
The Mott-insulating iron oxychalcogenides exhibit complex magnetic behaviour and we report here a neutron diffraction investigation into the magnetic ordering in $La_2O_2Fe_2OS_2$. This quaternary oxysulfide adopts the anti-$Sr_2MnO_2Mn_2Sb_2$-type structure (described by space group *I*4/*mmm*) and orders antiferromagnetically below $T_N$ = 105 K. We consider both its long-range magnetic structure and its magnetic microstructure, and the onset of magnetic order. It adopts the multi-k vector "2*k*" magnetic structure (*k* = (½ 0 ½) and *k* = (0 ½ ½) and has similarities with related iron oxychalcogenides, illustrating the robust nature of the "2*k*" magnetic structure.

## 1. Introduction

Mixed-anion systems, containing more than one kind of anion, often adopt anion-ordered structures with transition metal cations in unusual oxidation states and environments.[1-2] They therefore have the potential to exhibit interesting properties, including iron-based superconductivity in *Ln*FeAsO-related materials (*Ln* = lanthanide),[3-4] thermoelectricity in BiCuOSe[5] and wide-band gap semiconductivity in *Ln*CuO*Q* (*Q* = S, Se).[6-8] The anion ordering in these materials, resulting from the different sizes and characters of the oxide and pnictide or chalcogenide ions, often gives layered crystal structures with the oxide anions usually coordinating the "harder" cations in fluorite-like layers, whilst the "softer" transition metal cations are coordinated by the more covalent pnictide or chalcogenide anion. The ZrCuSiAs (or "1111") structure[9] (Figure 1a) adopted by *Ln*FeAsO, BiCuOSe and *Ln*CuO*Q* listed above is relatively simple but its building blocks, e.g. the fluorite-like oxide layers, or anti-fluorite-like transition metal layers, can be incorporated into more complex mixed anion materials to give new functional materials[10-18] and it's interesting to study the influence of the anion *Q* on the resulting properties and electronic structures.

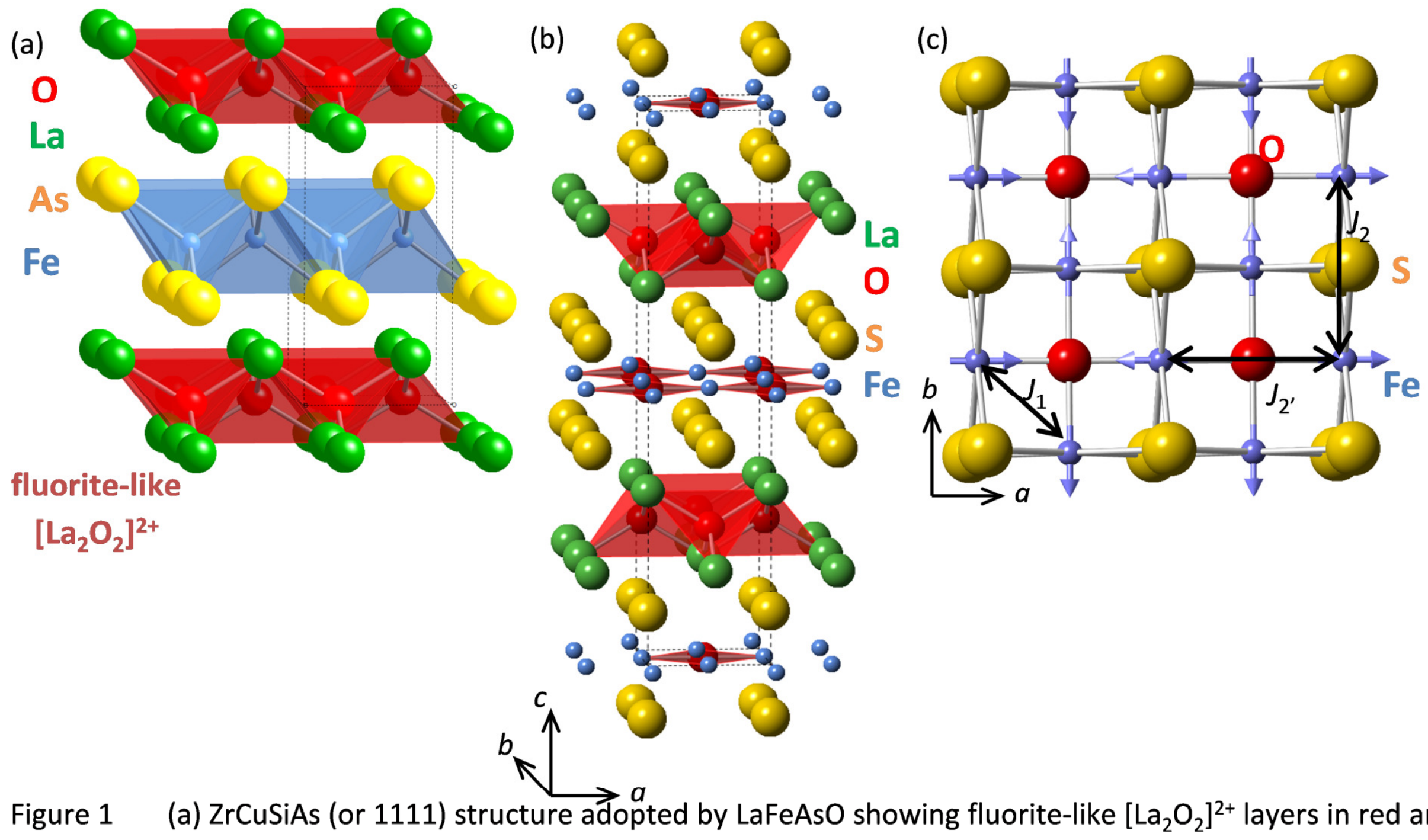

Figure 1 (a) ZrCuSiAs (or 1111) structure adopted by LaFeAsO showing fluorite-like $[La_2O_2]^{2+}$ layers in red and anti-fluorite-like $[Fe_2As_2]^{2-}$ layers in blue; (b) crystal structure of $La_2O_2Fe_2OS_2$ with La, Fe, O and S ions shown in green, blue, red and yellow, respectively (again, $[La_2O_2]^{2+}$ layers are in red) and (c) shows the in-plane 2*k* magnetic structure with $Fe^{2+}$ moments shown by blue arrows.

The iron oxychalcogenides $La_2O_2Fe_2OQ_2$ (*Q* = S, Se) were first reported in 1992[19] and adopt a body-centred tetragonal crystal structure (*I*4/*mmm*) consisting of fluorite-like $[La_2O_2]^{2+}$ layers and $[Fe_2O]^{2+}$ layers separated by $Q^{2-}$ anions (Figure 1b). (This structure can be described as an anti-$Sr_2MnO_2Mn_2Sb_2$-type structure[20] with the cation and anion sites swapped.)[1] Substitutions on the transition metal site (*M* = Mn, Fe, Co) and in the fluorite-like layers can be carried out and a number of members of this family have been reported.[2, 21-32] Their magnetism has been the focus of several studies because their magnetic order results from three competing exchange interactions: nearest-neighbour (nn) $J_1$ exchange interactions (either direct, or via 60 – 70° *M* – Se – *M* exchange interactions); next-nearest-neighbour (nnn) $J_2$ ~100° *M* – *Q* – *M* exchange and nnn $J_{2'}$ 180° *M* – O – *M* exchange (Figure 1b). The relative strengths of these exchange interactions changes with transition metal, with nn $J_1$ exchange dominating for the phases with the less electronegative Mn cation,[21, 25, 33] whilst AFM nnn $J_{2'}$ dominates for the dominates for the analogues with the more electronegative Co cation;[26, 28] the sign of the nnn $J_2$ *M* – *Q* – *M* exchange also changes with transition metal (ferromagnetic (FM) for *M* = Fe, Co[23, 29, 34-36] and antiferromagnetic (AFM) for *M* = Mn[33, 37]). For iron analogues (with iron of intermediate electronegativity), an AFM two *k*-vector magnetic structure (referred to here as the “2*k”* structure, described by the two perpendicular *k*-vectors $k_1$ = (½ 0 ½) and $k_2$ = (0 ½ ½)[30]) first proposed by Fuwa et al[27] has been reported for $Sr_2F_2Fe_2OS_2$[29] and for $La_2O_2Fe_2OSe_2$ (Figure 1c).[30] This non-collinear magnetic structure, with $Fe^{2+}$ moments directed along the Fe – O bonds[27, 38] (consistent with 2D-Ising-like character)[29-30] allows AFM nnn $J_{2'}$ Fe – O – Fe exchange and FM nnn $J_2$ Fe – *Q* – Fe exchange, while nn $Fe^{2+}$ moments are perpendicular to one another. The electronic structure of the iron-based “$Fe_2O$” systems has attracted much interest, with theoretical studies highlighting their Mott-insulating nature[34, 39-40] and possible relationship to the parent phases of the iron-based superconductors. Inelastic neutron scattering on the oxyselenide $La_2O_2Fe_2OSe_2$ suggests that the exchange interactions are weaker than previously thought, which may suggest some additional electron localisation which has not been fully explored theoretically.[30]

Although $La_2O_2Fe_2OS_2$ was first reported in 1992,[19] its magnetic structure has not been investigated by neutron powder diffraction (NPD) experiments. Liu et al recently investigated the role of the oxychalcogenide anion Q in the series $Nd_2O_2Fe_2OSe_{2-x}S_x$ and suggested that introducing $S^{2-}$ onto the chalcogenide site induced an enhanced FM component.[41] This prompted us to investigate the magnetic ordering and structure of closely-related $La_2O_2Fe_2OS_2$; we report here a structural and magnetic study using variable temperature NPD data that allows us to confirm its magnetic structure, and by comparison with related “$Fe_2O$” materials, to highlight the robust nature of this “2*k”* magnetic structure, and its role in giving a magnetic microstructure common to all “$Fe_2O$” materials studied.

## 2. Experimental

4.44 g of $La_2O_2Fe_2OS_2$ were prepared by the solid state reaction of $La_2O_3$ (Sigma-Aldrich, 99.99%), Fe (Alfa-Aesar, 99+%) and Se (Alfa-Aesar, 99+%). Stoichiometric quantities of these reagents were intimately ground together by hand using an agate pestle and mortar. The resulting grey powder was pressed into several 5 mm diameter pellets using a uniaxial press. These pellets were slowly heated in an evacuated, sealed silica tube to 400°C and held at this temperature for 12 hours, and then heated to 600°C and then 850°C held at this reaction temperature for 12 hours. The sample was then cooled to room temperature in the furnace. Preliminary structural characterisation was carried out using powder X-ray diffraction data collected on a Panalytical Empyrean diffractometer from 10° - 90° 2θ. The diffractometer was fitted with a germanium monochromator, an X'Celerator detector and an Oxford Cryosystems Phenix cryostat.

Neutron powder diffraction data were collected on the high flux diffractometer D20 at the ILL with neutron wavelength 2.41 °A. The powder was placed in an 10 mm diameter cylindrical vanadium can (to a height of 2.5 cm) and data were collected from 5-130° 2θ. A 40 minute scan was collected at 1.8 K and 10 minute scans were collected on warming at 2 K $min^{-1}$ to 168 K. Rietveld refinements were performed using TopasAcademic software.[42-43] The diffractometer zero point and neutron wavelength were refined using data collected at 160 K for which lattice parameters were known from XRPD analysis and were then fixed for subsequent refinements. A background was refined for each refinement, as well as unit cell parameters, atomic positions and a pseudo-Voight peak shape. TopasAcademic permits nuclear-only and magnetic-only phases to be included in refinements and the unit cell parameters of the magnetic phase were constrained to be integer multiples of those of the nuclear phase. The scale factor scales with the square of the unit cell volume; the scale factor for the nuclear phase was refined and that for the "2*k*" magnetic phase (with cell volume 8 times that of the nuclear phase) was constrained to be 0.015625 × that of the nuclear phase. The web-based ISODISTORT software was used to obtain a magnetic symmetry mode description of the magnetic structure;[44] magnetic symmetry modes were then refined corresponding to either the collinear magnetic structure, or the "2*k*" magnetic structure (see Section 3.2 below). Magnetic susceptibility data were measured using a Magnetic Properties Measurement System (MPMS, Quantum Design). Field-cooled and zero-field-cooled data were collected on warming from 2 K to 300 K at 5 K $min^{-1}$ in an applied magnetic field of 1000 Oe.

## 3. Results

### 3.1 Structural characterisation

Preliminary structural characterisation by Rietveld refinement using room temperature XRPD data indicated that our $La_2O_2Fe_2OS_2$ sample was of high purity, with room temperature lattice parameters $a$ = 4.04431(5) Å and $c$ = 17.8786(3) Å (see supplementary material), in good agreement with the crystal structure reported by Mayer et al.[19] NPD data were collected on a 4.44 g sample of $La_2O_2Fe_2OS_2$ on warming from ~1.8 K to ~168 K. NPD data collected at 168 K (above $T_N$) were consistent with the crystal structure reported by Mayer et al[19] and allowing Fe, S and O site occupancies to refine (whilst that of the La site was fixed at unity) gave site occupancies close to unity (Fe: 0.985(3); S: 0.985(8); O(1): 1.001(4) and O(2): 0.995(5)) (see supplementary material). The slight iron deficiency may indicate a very small amount of oxidation of $Fe^{2+}$ ions, but this stoichiometry is very close to ideal and the room temperature lattice parameters and property measurements (e.g. magnetic susceptibility measurements, see supplementary material) are consistent with those reported in the literature. Sequential Rietveld refinements using data collected on warming revealed a smooth increase in unit cell volume with temperature with expansion of ~0.08% along [100] and ~0.16% along [001] (Figure 2 and supplementary material).

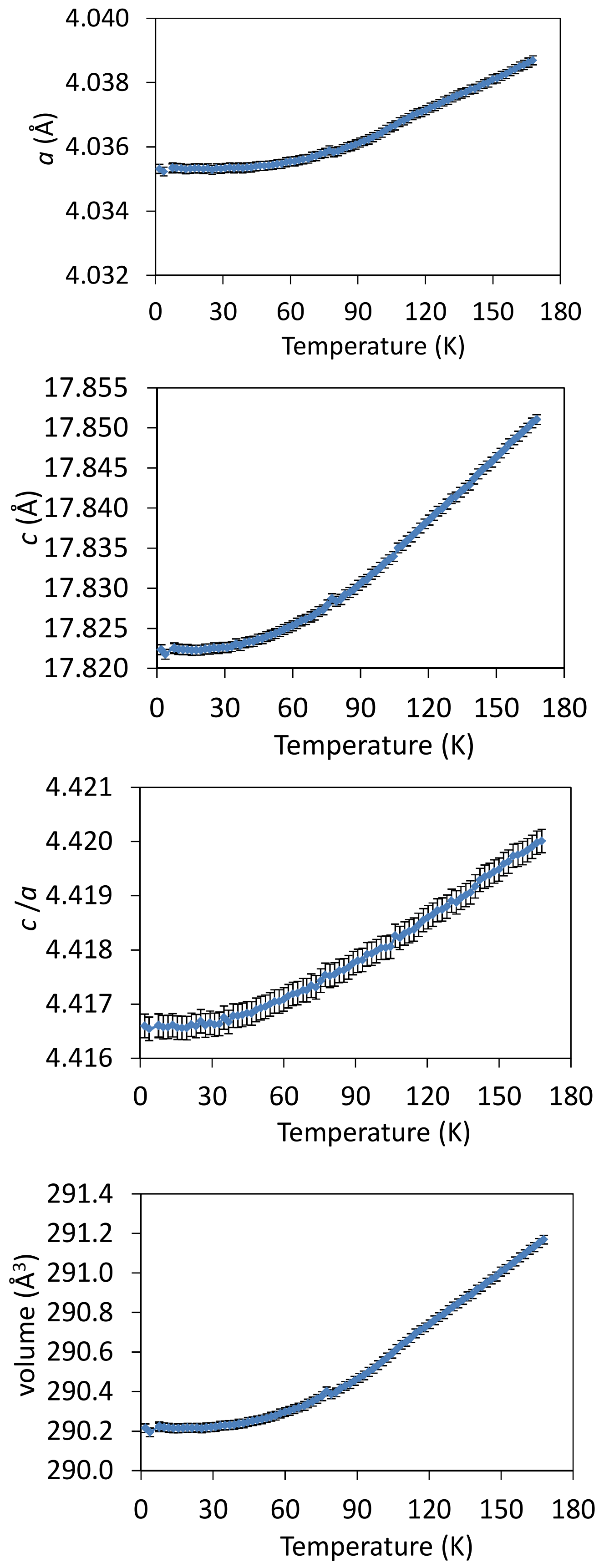

Figure 2 Unit cell parameters as a function of temperature for $La_2O_2Fe_2OS_2$ determined from sequential Rietveld refinements using NPD data.

### 3.2 Magnetic ordering

A broad asymmetric Warren-like peak[45] centred at ~39° 2θ was observed in data collected immediately above $T_N$ (from 106.5 K) which decreased in intensity rapidly and could not be detected above 116 K. This peak, centred around the position of the most intense magnetic Bragg reflection (which appears below $T_N$) is characteristic of two-dimensional short-ranged magnetic order,[45] similar to the magnetic Warren peak observed for other $Ln_2O_2Fe_2OSe_2$ analogues (*Ln* = La, Ce, Pr).[30-32] Fitting with a Warren function gave a correlation length of ~100 Å (~25 times in the in-plane lattice parameter) at 106.5 K (immediately above $T_N$) which decreases very rapidly (Figure 3 and supplementary material).

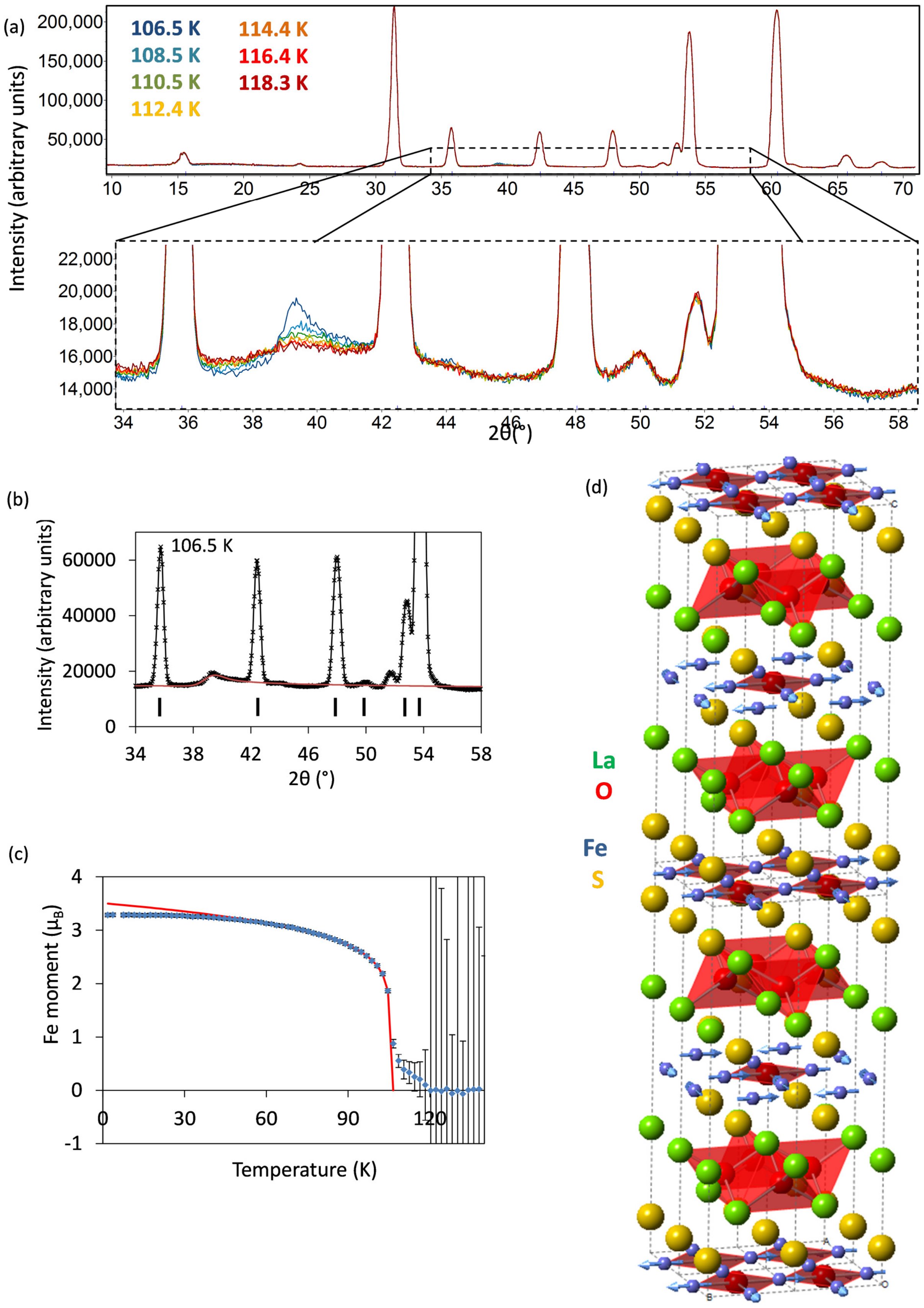

Figure 3 [colour online] (a) NPD data collected for $La_2O_2Fe_2OS_2$ close to $T_N$ showing (above) 10 – 70° 2θ data and (below, enlarged) the narrow 2θ region in which the Warren peak is observed and (b) shows the fit to this Warren peak in 106.5 K data to a model suggesting an in-plane magnetic correlation length of ~100 Å (observed data points in black; calculated Warren profile + background shown by red line; peak positions for the nuclear phase are shown by black tick marks); (c) shows the evolution of Fe2+ magnetic moment on cooling with data points in blue and solid red showing as a guide to the eye showing critical behaviour for a 2D-Ising system (with critical exponent β = 0.132(1), $T_N$ = 105.48(3) K and $M_0$ = 3.508(7) $\mu_B$); (d) shows nuclear and magnetic structure at 1.8 K for $La_2O_2Fe_2OS_2$ with La, Fe, O and S ions shown in green, blue, red and yellow, respectively, and $Fe^{2+}$ moments shown by blue arrows.

Additional Bragg reflections are observed in data collected for $T \leq 104.5$ K which increase smoothly in intensity with decreasing temperature. The web-based ISODISTORT software[44] was used, in conjunction with TopasAcademic refinement software[42-43] to explore possible magnetic structures.

Indexing these additional reflections suggested a magnetic unit cell described by *k* vector *k* = (0 ½ ½) as proposed for $La_2O_2Fe_2OSe_2$ and collinear magnetic order with a high degree of frustration.[24] However, similar magnetic Bragg scattering has since been observed for related iron oxychalcogenides and found to be fitted equally well by a non-collinear two-*k*-vector magnetic structure (the "2*k*" magnetic structure).[29-30]

The magnetic Bragg reflections observed for $La_2O_2Fe_2OS_2$ were found to be consistent with the "2*k*" magnetic structure proposed by Fuwa et al[27] for $Nd_2O_2Fe_2OSe_2$ and by Zhao et al for $Sr_2F_2Fe_2OS_2$.[29] (Equally good fits could be obtained using a collinear model (as described by Free et al[24]), but given the narrow temperature-range for which the Warren-peak is observed (suggesting low levels of frustration) and the observed 2D-Ising-like onset of $Fe^{2+}$ magnetic moment on cooling (Figure 3c), the "2*k*" magnetic ordering seems more likely for this oxysulfide, consistent with closely-related materials. [29-32] )

Initially, our Rietveld model (including the nuclear phase and a magnetic-only phase describing "2*k*" magnetic order on the $Fe^{2+}$ sublattice) gave a poor fit to the data (Figure 4a) due to significant anisotropic broadening of the magnetic Bragg reflections. This broadening was fitted with a model describing antiphase boundaries (e.g. stacking faults)[46] perpendicular to the *c* axis in the magnetic structure which gave a good fit to the data (Figures 4b, c). The magnetic correlation length along *c*, $\xi_c$ is ~51(2) Å at 2 K and changes little with temperature ($\xi_c$ = 55(6) at 103 K). We note at two additional weak reflections (at 24° 2θ and at 52° 2θ) are observed in our NPD data at all temperatures (Figures 3, 4 and supplementary materials) and are thought to be due to a small amount of an unidentified impurity phase, and are not thought to relate to the magnetic behaviour of $La_2O_2Fe_2OS_2$, the focus of this study.

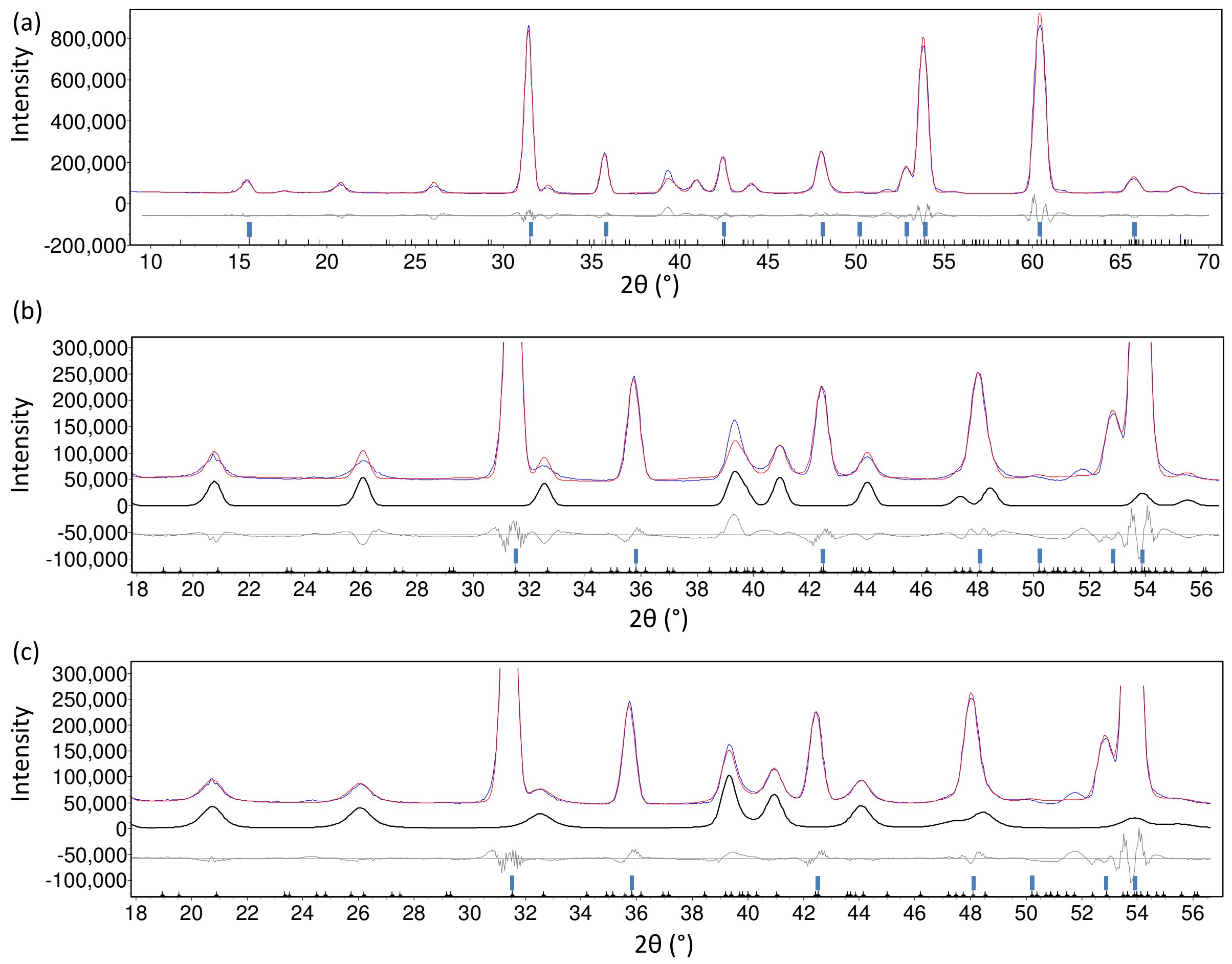

Figure 4 Rietveld refinement profiles with "2*k*" magnetic model at 1.8 K showing (a) wide 2θ range with reflections for nuclear and magnetic phases shown by blue and black ticks, respectively, (b) narrow 2θ range highlighting fit to magnetic reflections with the same peak shape for both nuclear and magnetic phases and (c) showing improved fit to magnetic reflections by the "2*k*" magnetic model with stacking faults. Magnetic scattering is highlighted by solid black line in panels (b) and (c), whilst the observed, calculated and difference lines are shown in blue, red and grey, respectively.

Sequential NPD Rietveld refinements indicate a smooth increase in $Fe^{2+}$ moment on cooling (Figure 3c and supplementary material). Fitting this magnetic order parameter by models for critical behaviour suggests similar 2D-Ising-like character around $T_N$ before long-ranged three-dimensional order develops, as observed for other "$Fe_2O$" materials[30-32] and consistent with "2*k*" magnetic ordering with $Fe^{2+}$ moments directed along the Fe – O bonds. The low temperature (1.8 K data) were fitted well by a model containing the *I*4/*mmm* nuclear phase and a magnetic-only phase describing "2*k*" magnetic order on the $Fe^{2+}$ sublattice (with stacking faults in the magnetic structure as described above) and an ordered moment of 3.30(4) $\mu_B$ on $Fe^{2+}$ sites. This is in good agreement with similar models for other "$Fe_2O$" materials including $Sr_2F_2Fe_2OS_2$ ($Fe^{2+}$ moment = 3.3(1) $\mu_B$)[29] and $La_2O_2Fe_2OSe_2$ ($Fe^{2+}$ moment = 3.50(5) $\mu_B$).[30] Full details from the 1.8 K refinement and selected bond lengths are given in Tables 1 and 2 and the final structure (nuclear and magnetic) is illustrated in Figure 3d.

Table 1 Details from Rietveld refinement using 1.8 K NPD data for $La_2O_2Fe_2OS_2$. The refinement was carried out with the nuclear structure described by space group *I*4/*mmm* with *a* = 4.0353(1) Å and *c* = 17.8237(9) Å, and "2*k*" magnetic ordering on the $Fe^{2+}$ sublattice (as described above) with magnetic correlation length $\xi_c$ = 51(2) Å; $R_{wp}$ = 5.680% and $R_p$ = 3.961%, $R_B$ = 0.89% (nuclear phase) and $R_B$ = 1.80% (magnetic phase).

| **Atom** | **Site** | ***x*** | ***y*** | ***z*** | **$U_{iso}$ × 100 (Å$^2$)** | **$Fe^{2+}$ moment ($\mu_B$)** |
|---|---|---|---|---|---|---|
| La | 4*e* | 0.5 | 0.5 | 0.1803(1) | 0.9(2) | |
| Fe | 4*c* | 0.5 | 0 | 0 | 0.9(2) | 3.30(4) |
| S | 4*e* | 0 | 0 | 0.0934(5) | 0.9(2) | |
| O(1) | 4*d* | 0.5 | 0 | 0.25 | 0.9(2) | |
| O(2) | 2*b* | 0.5 | 0.5 | 0 | 0.9(2) | |

Table 2 Selected bond lengths and angles from Rietveld refinement using 1.8 K NPD data for $La_2O_2Fe_2OS_2$.

| | |
|---|---|
| **Bond lengths (Å)** | |
| La – O(1) | 4 × 2.370(1) |
| La – S | 4 × 3.214(2) |
| Fe – Fe | 4 × 2.85341(9) |
| Fe – O(2) | 2 × 2.01766(6) |
| Fe – S | 4 × 2.616(6) |
| **Bond angles (°)** | |
| Fe – O – Fe | 180 |
| Fe – S – Fe(1) | 66.1(1) |
| Fe – S – Fe(2) | 100.9(3) |

## 4. Discussion

The crystal structure of $La_2O_2Fe_2OS_2$ is very similar to that of the oxide-fluoride-sulfide $Sr_2F_2Fe_2OS_2$.[23] The slightly larger $Sr^{2+}$ cation (eight-coordinate ionic radii are 1.26 Å for $Sr^{2+}$ and 1.160 Å for $La^{3+}$)[47] in the fluorite-like $[A_2X_2]^{2+}$ layers of the latter give a slightly larger separation of the "$Fe_2O$" layers, but this has very little effect on the AFM ordering temperature ($T_N$ = 106(2) K for $Sr_2F_2Fe_2OS_2$;[23, 29] 105(1) K for $La_2O_2Fe_2OS_2$ here).

We note that the change in *c* lattice parameter with temperature for this oxysulfide $La_2O_2Fe_2OS_2$ does not show as marked a discontinuity at $T_N$ as the analogous oxyselenide $La_2O_2Fe_2OSe_2$ (see supplementary material for comparison).[24] The more rapid decrease in *c* lattice parameter below $T_N$ for oxyselenides $Ln_2O_2Fe_2OSe_2$ was ascribed to magnetostrictive effects[24, 31-32] and it is noticeably less pronounced in the oxysulfide $La_2O_2Fe_2OS_2$ with its shorter *c* lattice parameter. This is consistent with the trend observed across $Ln_2O_2M_2OQ_2$ systems (*Ln* = lanthanide ion; *M* = Mn, Fe, Co; *Q* = S, Se) so far, in which the more rapid decrease in *c* below $T_N$ occurs for systems with larger separation between subsequent magnetic "$M_2O$" layers.[25, 31-32]

The observation of a Warren peak immediately above $T_N$ suggests some two-dimensional short-ranged magnetic order before three-dimensional magnetic order develops below $T_N$. This is consistent with the broad maximum in magnetic susceptibility measurements (see Mayer et al[19] and supplementary material). The observation of the Warren peak over such a narrow temperature range above $T_N$ (~ 10 K here for $La_2O_2Fe_2OS_2$; ~14 K for $La_2O_2Fe_2OSe_2$;[30] ~140 K for $La_2O_2Mn_2OSe_2$[21]) reflects the low degree of magnetic frustration expected for the "2*k*" magnetic structure,[30] with both FM $J_2$ (Fe – Se – Fe) and AFM $J_{2'}$ (Fe – O – Fe) nnn interactions satisfied. We note that the occurrence of this multi-*k* non-collinear magnetic structure rather than a single-*k* structure implies the presence of higher-order terms which couple the two orthogonal *k*-vectors and maintain the $C_4$ symmetry of the nuclear crystal structure,[30] whilst the collinear structure[24] breaks this tetragonal symmetry.

NPD data for several "$Fe_2O$" materials suggest stacking faults in the magnetic structure and it's interesting that this "magnetic microstructure" observation is common to all "$Fe_2O$" materials studied, but hasn't been reported for manganese or cobalt analogues which have sharp magnetic Bragg reflections. The stacking faults in the $Fe^{2+}$ magnetic structure are likely to result from the body-centred crystal structure, which means that consecutive "$Fe_2O$" layers are offset from one another by (½ ½ 0) and therefore the O(2) site alternates between (0 0 *z*) and (½ ½ *z*) in successive layers. Although the $Fe^{2+}$ sites are coincident in consecutive layers, $Fe^{2+}$ moments (directed along Fe – O bonds) in consecutive layers are perpendicular to one another. Flipping the direction of moments in one layer (by 180°) relative to the layer below is therefore likely to be a relatively low energy defect. This would explain the presence of stacking faults in "$Fe_2O$" materials and their absence in manganese analogues with their collinear magnetic structure. We note that the magnetic correlation length along *c*, $\xi_c$, is slightly longer in this oxysulfide (~51(2) Å at 2 K with only 8.9119(9) Å between magnetic "$Fe_2O$" layers) than in the analogous oxyselenide $La_2O_2Fe_2OSe_2$ (45(3) Å at 2 K with 9.258(5) Å between layers), presumably due to the slightly closer "$Fe_2O$" layers in the oxysulfide.

The magnetic behaviour of $La_2O_2Fe_2OS_2$ is extremely similar to that of $Sr_2F_2Fe_2OS_2$[29] as well as to that of the analogous oxyselenide $La_2O_2Fe_2OSe_2$, although the exchange interactions in the oxysulfides are slightly stronger resulting in a slightly higher $T_N$ than for the oxyselenides (presumably due to the shorter bond distances and better overlap in the oxysulfides). This suggests that the relative strengths of these exchange interactions, and the $Fe^{2+}$ anisotropy are unchanged. Liu et al suggest that as selenide ions are replaced by smaller sulfide ions in $Nd_2O_2Fe_2OSe_{2-x}S_x$, a FM component is induced as the relative strengths of the exchange interactions are modified by chemical pressure.[41] Comparing room temperature crystal structures for various "$Fe_2O$" materials indicates that the *Q* – Fe – *Q* angles are similar for both $Nd_2O_2Fe_2OSe_{1.6}S_{0.4}$ and for $Nd_2O_2Fe_2OSe_2$ ($\alpha_1$ is 95.88(2)° for $Nd_2O_2Fe_2OSe_{1.6}S_{0.4}$[41] and 95.72(2)° for $Nd_2O_2Fe_2OSe_2$[31]). While Fe – O and Fe – Fe distances are comparably shorter for $Nd_2O_2Fe_2OSe_{1.6}S_{0.4}$ than for $Nd_2O_2Fe_2OSe_2$,[31] $La_2O_2Fe_2OS_2$ (here) and $Sr_2F_2Fe_2OS_2$,[23] it's interesting that the room-temperature Fe – *Q* bond length for $Nd_2O_2Fe_2OSe_{1.6}S_{0.4}$ (2.7026(3) Å)[41] is intermediate between those reported for $Nd_2O_2Fe_2OSe_2$ (2.7154(4) Å)[31] and for $Sr_2F_2Fe_2OS_2$ (2.633 Å).[23] The similar magnetic behaviour observed for $La_2O_2Fe_2OS_2$ (here), $Sr_2F_2Fe_2OS_2$[29] and $Ln_2O_2Fe_2OSe_2$[30-31] (including critical behaviour, long-range magnetic ordering, degree of frustration and magnetic microstructure) illustrates the robust nature of the "2*k*" magnetic order in "$Fe_2O$" materials, little changed with anion *Q* or small variations in crystal structure. However, it raises the question as to whether the ~300 K FM component observed in magnetic susceptibility data for $Nd_2O_2Fe_2OSe_{2-x}S_x$[41] is intrinsic to the material or might arise from trace amounts of a ferromagnetic impurity. Inelastic neutron scattering studies to determine the strengths of the exchange interactions in these oxysulfides would be interesting and allow comparison with the Mott-insulating oxyselenides.

## 5. Conclusions

The magnetic behaviour in $La_2O_2Fe_2OS_2$ reported here is extremely similar to that of other "$Fe_2O$" materials which illustrates the robust nature of the "2*k*" magnetic structure. This magnetic ordering is dominated by nnn $J_2$ and $J_{2'}$ interactions and the $Fe^{2+}$ magnetic anisotropy and is relatively independent of variations in unit cell size, the chalcogenide $Q^{2-}$ and structural distortions.[32] The onset of magnetic order in this oxysulfide is very sudden with two-dimensional short-ranged order developing in a narrow temperature range immediately above $T_N$, in contrast to the more three-dimensional like character of the manganese analogue.[21] We've shown that the stacking faults in the magnetic structure of these "$Fe_2O$" materials are a consequence of the 2D-Ising nature of the two-*k* vector magnetic order.

**Acknowledgements**

We are grateful to Mr Ben Coles for assistance collecting NPD data and to Dr Andrew Wills for helpful discussions.

**Graphical Abstract**

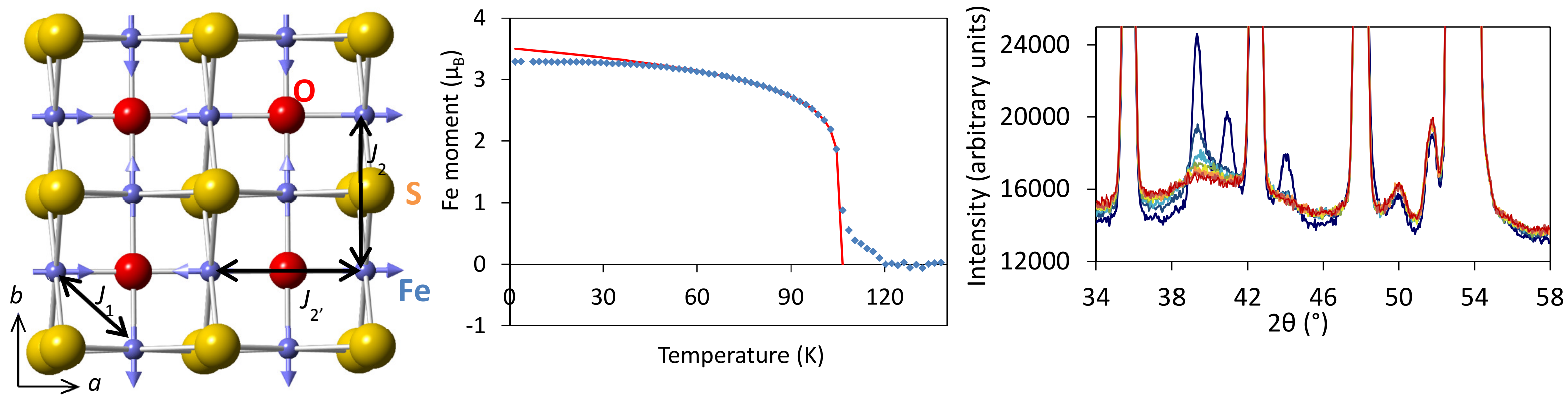

**Supplementary material**

SM1 Rietveld refinement profiles and details using room temperature XRPD data.

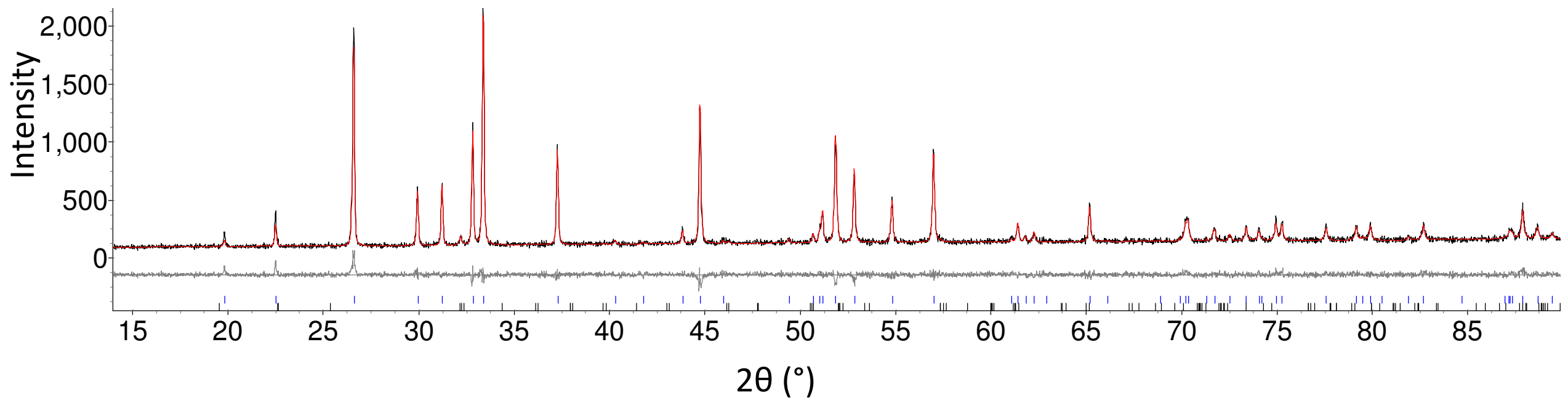

Figure SM1 Rietveld refinement profiles for $La_2O_2Fe_2OS_2$ using room temperature XRPD data showing observed (black), calculated (red) and difference (grey) profiles with tick marks showing the positions of the main phase (upper blue; 98.9%) and $LaFeO_3$ impurity (lower black, 1.1%). $R_{wp}$ = 8.81% and $R_p$ = 6.98%.

SM2 168 K NPD refinement

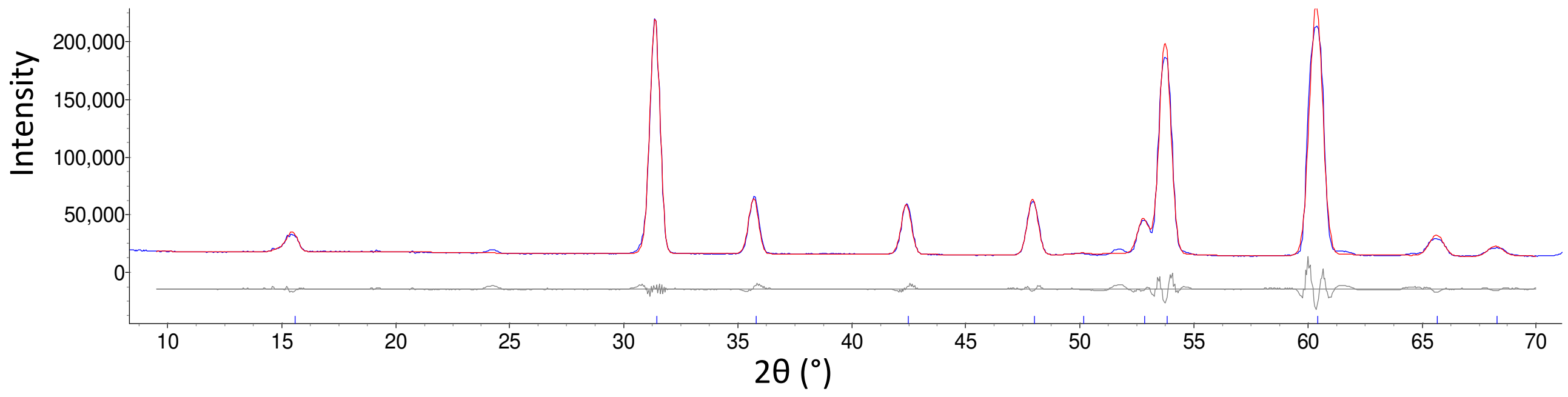

Figure SM2 Rietveld refinement profiles for $La_2O_2Fe_2OS_2$ using 168 K NPD data showing observed (blue), calculated (red) and difference (grey) profiles with tick marks showing the peak positions for the tetragonal nuclear structure.

Table1 SM2 Details from Rietveld refinement using 168 K NPD data for $La_2O_2Fe_2OS_2$. The refinement was carried out with the nuclear structure described by space group *I*4/*mmm* with *a* = 4.0387(1) Å and *c* = 17.8522(9) Å; $R_{wp}$ = 5.367% and $R_p$ = 3.534%, $R_B$ = 0.99% (nuclear phase).

| **Atom** | **Site** | ***x*** | ***y*** | *z* | $U_{iso}$ **× 100 (Å$^2$)** |
|---|---|---|---|---|---|
| La | 4*e* | 0.5 | 0.5 | 0.1805(1) | 1.2(2) |
| Fe | 4*c* | 0.5 | 0 | 0 | 1.2(2) |
| S | 4*e* | 0 | 0 | 0.0932(5) | 1.2(2) |
| O(1) | 4*d* | 0.5 | 0 | 0.25 | 1.2(2) |
| O(2) | 2*b* | 0.5 | 0.5 | 0 | 1.2(2) |

SM3 Magnetic susceptibility measurements for $La_2O_2Fe_2OS_2$

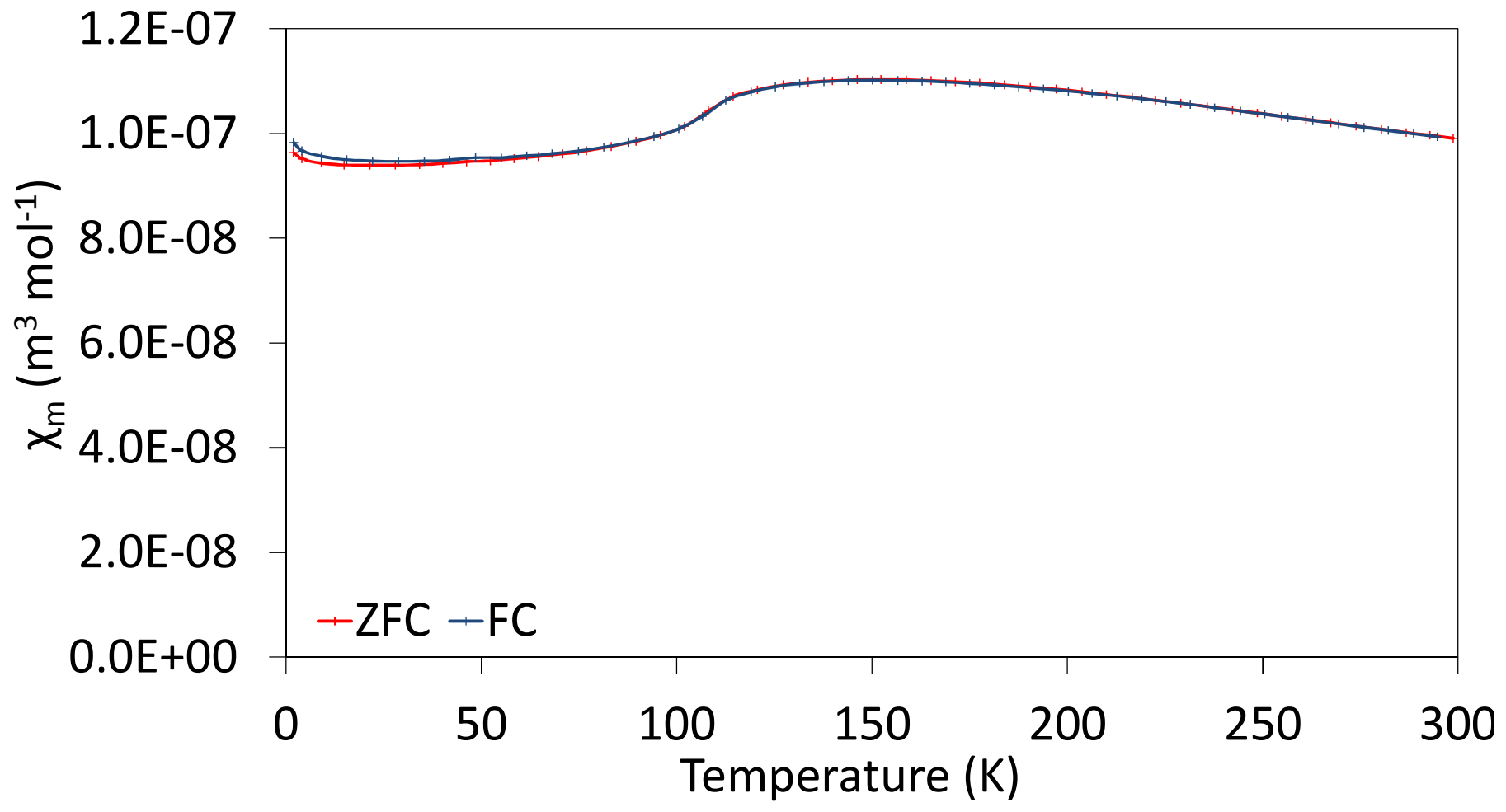

SM4 NPD data collected on warming from 1.8 K to 168 K for $La_2O_2Fe_2OS_2$; ticks for the nuclear and magnetic phases are shown in black and blue, respectively.

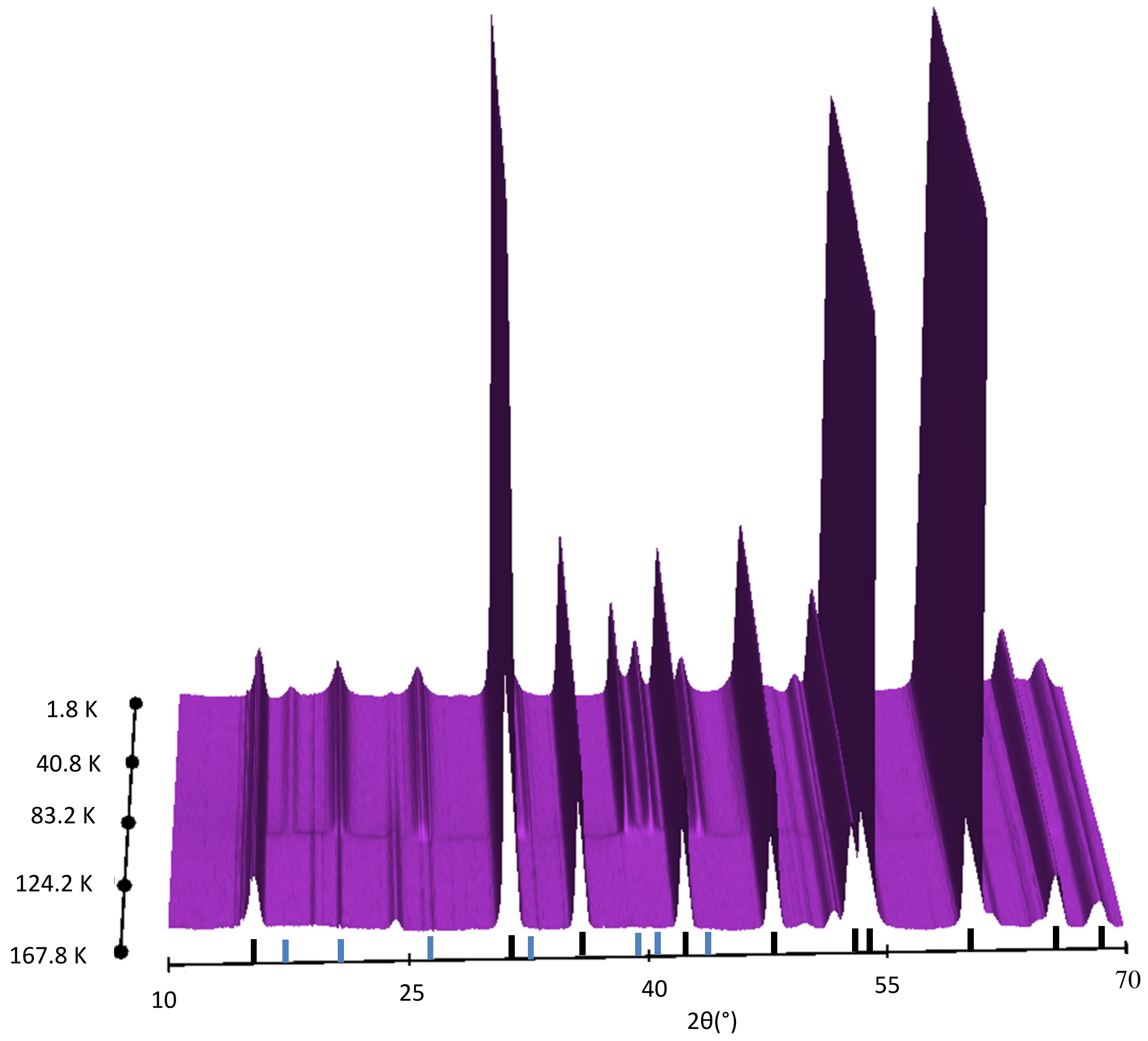

SM5 Results from sequential Rietveld refinements using NPD data collected for $La_2O_2Fe_2OS_2$ on warming.

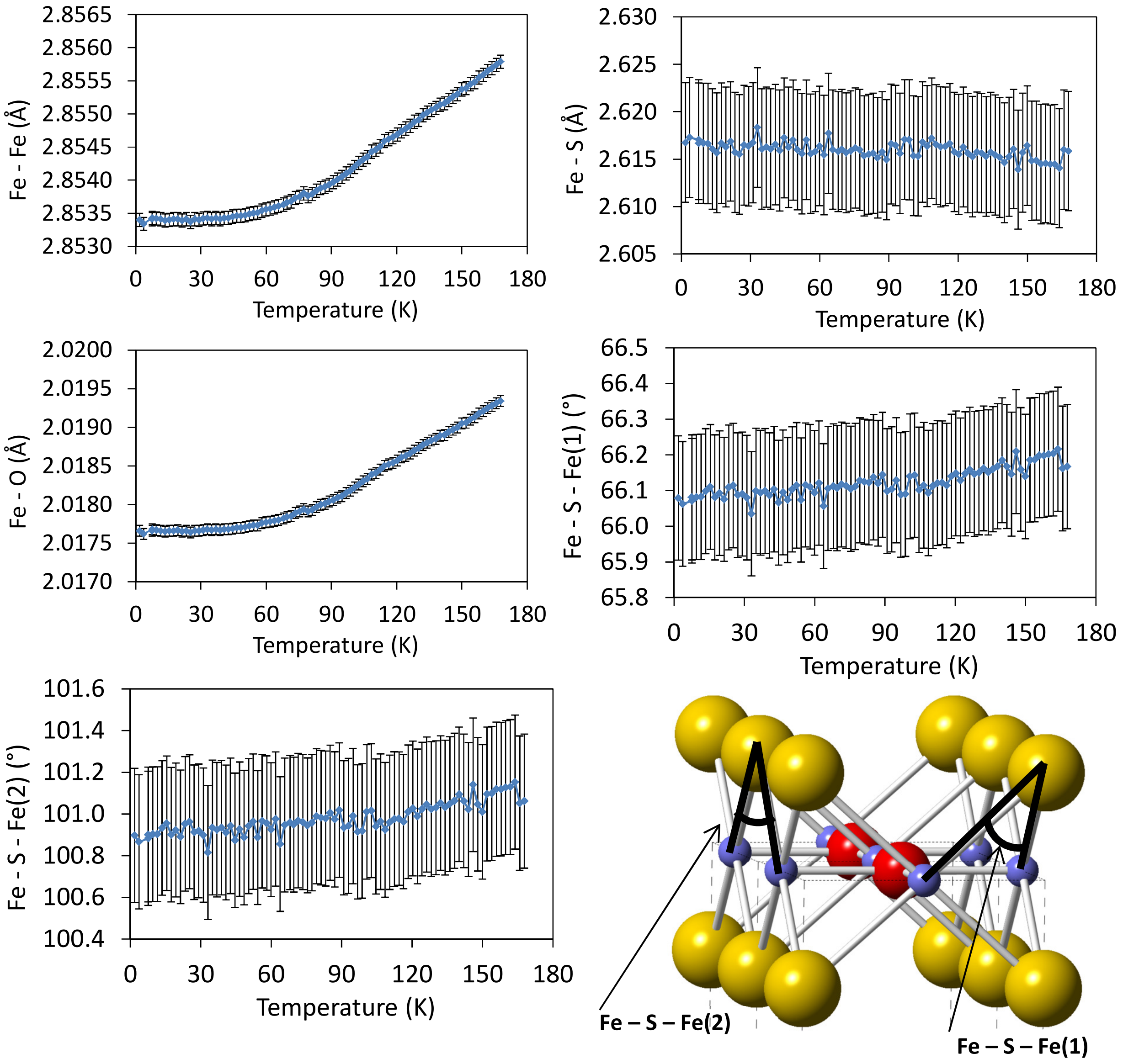

Figure SM5 Selected bond lengths and angles for $La_2O_2Fe_2OS_2$ as a function of temperature, determined from sequential Rietveld refinements using NPD data. We notes that the relatively large esds on some bond lengths reflect the very small change in these values in the temperature range studied, and the low Q-range of our data (making it relatively insensitive to subtle structural changes).

SM6 Comparison of unit cell parameters on cooling for $La_2O_2Fe_2OS_2$ and $La_2O_2Fe_2OSe_2$.

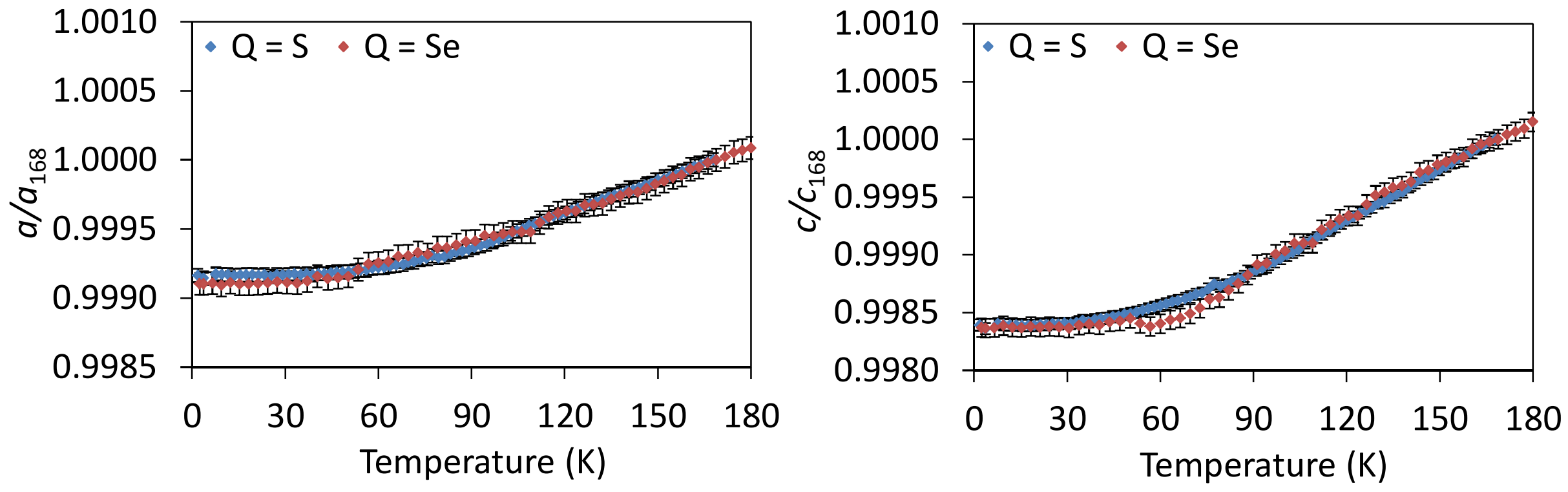

Figure SM6 Comparison of normalised unit cell parameters for $La_2O_2Fe_2OS_2$ (*Q* = S, results reported here) and $La_2O_2Fe_2OSe_2$ (*Q* = Se, from reference 30). Unit cell parameters on cooling were normalised to their values at 168 K (above $T_N$) for both materials to aid comparison.

SM7 Comparison of evolution of $Fe^{2+}$ moment on cooling for $La_2O_2Fe_2OQ_2$ (*Q* = S, Se)

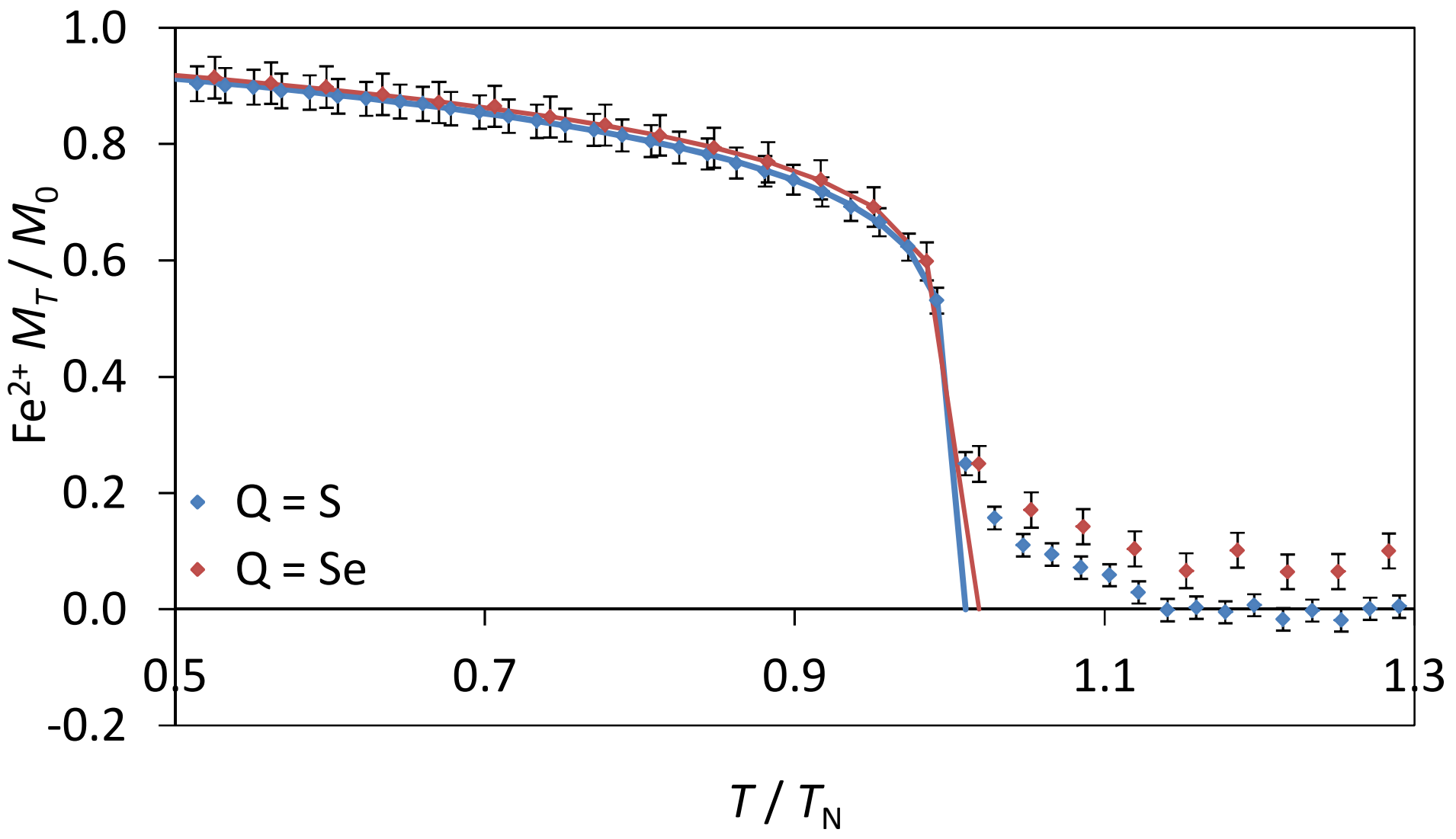

Figure SM7 Comparison of evolution of $Fe^{2+}$ moment on cooling for $La_2O_2Fe_2OS_2$ (*Q* = S; blue; results reported here) and $La_2O_2Fe_2OSe_2$ (*Q* = Se; red; reference 30). Data points are shown with error bars, and solid lines are fits to critical behaviour with parameters:

*Q* = S: $M_0$ = 3.508(7) $\mu_B$, $T_N$ = 105.48(3) K, β = 0.132 (1).

*Q* = Se: $M_0$ = 3.701(8) $\mu_B$, $T_N$ = 89.50(3) K, β = 0.122(1).